\documentclass[a4paper,fleqn,usenatbib]{mnras}
\usepackage{newtxtext,newtxmath}

\usepackage[T1]{fontenc}
\usepackage{ae,aecompl}

\usepackage{graphicx}	
\usepackage{amsmath}	
\usepackage{amssymb}	
\usepackage{threeparttable}
\usepackage{hyperref}
\usepackage{rotating}
\usepackage{subfigure}
\title[Host Galaxy Properties of CL-AGNs]{Host Galaxy Properties of Changing-look AGN Revealed in the MaNGA Survey}

\author[X. L. Yu et al.]{
Xiaoling Yu,$^{1,2}$
Yong Shi,$^{1,2}$\thanks{E-mail: yong@nju.edu.cn}
Yanmei Chen,$^{1,2}$
Jianhang Chen,$^{1,2}$
Songlin Li$^{1,2}$
\newauthor
Longji Bing$^{1,2}$
Junqiang Ge,$^{3}$
Rogemar A. Riffel,$^{4,5}$
and Rog\'erio Riffel$^{5,6}$
\\
$^{1}$School of Astronomy and Space Science, Nanjing University, Nanjing 210093, China\\
$^{2}$Key Laboratory of Modern Astronomy and Astrophysics (Nanjing University), Ministry of Education, Nanjing 210093, China\\
$^{3}$Chinese Academy of Sciences National Astronomical Observatories of China 20A Datun Road, Chaoyang District, Beijing 100012, China\\
$^{4}$Departamento de F\'isica, CCNE, Universidade Federal de Santa Maria, 97105-900, Santa Maria, RS, Brazil\\
$^{5}$Laborat\'orio Interinstitucional de e-Astronomia - LIneA, Rua Gal. Jos\'e Cristino 77, Rio de Janeiro, RJ - 20921-400, Brazil\\
$^{6}$Instituto de F\'isica, Universidade Federal do Rio Grande do Sul, Campus do Vale, Porto Alegre, RS, Brasil, 91501-970
}

\date{Accepted XXX. Received YYY; in original form ZZZ}

\pubyear{2019}

\begin{document}
\label{firstpage}
\pagerange{\pageref{firstpage}--\pageref{lastpage}}
\maketitle

\begin{abstract}
Changing-look Active Galactic Nuclei (CL-AGNs) are a subset of AGNs in which the broad Balmer emission lines appear or disappear within a few years. We use the Mapping Nearby Galaxies at Apache Point Observatory (MaNGA) survey to identify five CL-AGNs. The 2-D photometric and kinematic maps reveal common features as well as some unusual properties of CL-AGN hosts as compared to the AGN hosts in general. All MaNGA CL-AGNs reside in the star-forming main sequence, similar to MaNGA non-changing-look AGNs (NCL-AGNs). The $80\% \pm 16\%$ of our CL-AGNs do possess pseudo-bulge features, and follow the overall NCL-AGNs $M_{BH}-\sigma_{*}$ relationship. The kinematic measurements indicate that they have similar distributions in the plane of angular momentum versus galaxy ellipticity. MaNGA CL-AGNs however show a higher, but not statistically significant ($20\% \pm 16\%$) fraction of counter-rotating features compared to that ($1.84\% \pm 0.61\%$) in general star-formation population. In addition, MaNGA CL-AGNs favor more face-on (axis ratio $>$ 0.7) than that of Type I NCL-AGNs. These results suggest that host galaxies could play a role in the CL-AGN phenomenon.

\end{abstract} 


\section{Introduction}
Active galactic nuclei (AGNs) are empirically classified into type 1 and type 2 according to their emission line features. Type 1 AGNs possess both broad and narrow emission lines, while only narrow emission lines are seen in type 2 AGNs. The unification model \citep{Antonucci1993, Urry95, Netzer15} states that type 1 AGNs are viewed face on so that broad emission line regions (BLRs) are seen directly, while type 2 AGNs are viewed edge-on with their BLRs blocked by the dusty torus. Despite the success of the unification model, there are proposals that some type 2 AGNs may lack intrinsic broad lines because of their low accretion rate \citep{Shi10,Pons14}.

It has been known for decades that some AGNs have changed from type 1 to type 2 or vice versa \citep{Khachikian71, Penston84}. This type of AGNs, called changing-look AGNs (CL-AGNs), experience the appearance or disappearance of broad Balmer emission lines over a timescale of a few years. The number of such AGN has increased rapidly in the past few years due to large area spectroscopic surveys \citep[e.g.,][]{Denney14,LaMassa15,Runnoe16,MacLeod16,Ruan16,Parker16,Yang18,MacLeod19}. Some CL-AGNs have been observed to change types more than once: Mrk 1018 changed from type 1.9 to type 1 and later returned back to type 1.9 \citep{Cohen86, McElroy16}, and Mrk 590 has changed from type 1.5 to type 1 then to type 1.9, as seen from monitoring over a period of more than 40 years \citep{Denney14}. Some objects display rapid state change, e.g., it requires just a few months for 1ES 1927+654 to transform from type 2 to type 1 \citep{Trakhtenbrot19}. The physical mechanisms of CL-AGNs are still not well understood. Proposed mechanisms include: 1) variable obscuration where the BLR is obscured by crossing clouds \citep[e.g.,][]{Goodrich89,Elitzur12}; 2) variable accretion rates where AGNs follow sequence from type 1 to type 2 or vice versa \citep[e.g.,][]{Elitzur14}; and 3) tidal disruption events (TDEs) where a star disrupted by the supermassive black hole results in AGN type change \citep{Merloni15}.

In recent decades, more and more works reveal some links between small-scale supermassive black holes (SMBHs) accretion activities and large-scale host galaxy properties. \citet{Dunlop03} studied morphologies of AGN hosts, and pointed out that the hosts of both radio-loud and radio-quiet AGN are virtually all massive ellipticals. \citet{Jahnke04} found that the UV colors of all host galaxies are substantially bluer than expected from an old population of stars with formation redshift less than 5. Researchers also studied the interstellar medium of host galaxies \citep[e.g.,][]{xia12,Petric15}. They found that there is a correlation between the AGN-associated bolometric luminosities and the CO line luminosities, suggesting that star formation and AGNs draw from the same reservoir of gas. Studies also investigate the star formation behavior of the host galaxies \citep[e.g.,][]{Shi07,Shi14,Xu15}, and pointed out that far-infrared emission of AGNs can be dominated by either star formation or nuclear emission. These studies imply that the host galaxy properties are the key point to understanding the environment where the black hole grows, and provide more insights into the circumstances influencing co-evolution \citep[e.g.,][]{Zhang16}.

There is a tight correlation between the black hole mass ($M_{BH}$) and stellar velocity dispersion of the bulge component ($\sigma_{*}$) for nearby galaxies \citep[e.g.][]{Gultekin09,Xiao11,Kormendy13}. This relation suggests a strong connection between AGNs and their host galaxies. CL-AGNs offer a unique opportunity to study this relationship. \citet{Gezari17} estimated the $M_{BH}$ of one CL-AGN through the broad Balmer emission lines when the quasar transformed to type 1. They reported that the $M_{BH}$ is in good agreement with the estimate through the $M_{BH}-\sigma_{*}$ relation. The ``turning-off'' CL-AGNs provide excellent cases to investigate the host galaxies properties, as the contamination from nuclear radiation is greatly reduced \citep[e.g.,][]{LaMassa15,Yang18,Charlton19,Ruan19}. \citet{Raimundo19} used optical and near-infrared integral field unit (IFU) spectroscopy to study the host galaxy properties of Mrk 590, a CL-AGN that has been subject of a 40-year multi-wavelength monitoring campaign. This is the first IFU study of the host galaxy and central AGN properties. They found that the ionized gas has a complex shape and a complex relationship between gas inflow and outflow in the nucleus of Mrk 590.

In this work, we attempt to expand the optical IFU studies of CL-AGN by using the MaNGA survey. Section 2 describes the MaNGA survey and the sample selection criteria. Section 3 presents data analysis and results. We summarize our main conclusions in section 4. We adopt the cosmological parameters $H_0$ = 70 $\rm{km\ s^{-1}\ Mpc^{-1}}$, $\Omega_m$ = 0.3, $\Omega_{\Lambda}$ = 0.7 throughout this paper.

\section{Data}
\subsection{The MaNGA survey}
The MaNGA survey \citep{Bundy15} is an integral field spectroscopic program, which is one of three core programs in the fourth generation of the Sloan Digital Sky Survey (SDSS-IV) \citep{Blanton17}. This survey is operational from July 1, 2014 to June 30, 2020. The aim is targeting about 10,000 nearby galaxies within a redshift range of 0.01 $\sim$ 0.15 \citep{Drory15, Law15, Yan16a,Wake17}, which are observed by the 2.5-m Sloan Foundation Telescope\citep{Gunn06}. The wavelength coverage is from 3600 $\rm\AA$ to 10300 $\rm\AA$ with the spectral resolution R $\sim$ 2,000 \citep{Smee13}. Two-thirds of the targets will be covered out to 1.5 effective radii (Re), the remaining one-third out to 2.5 Re \citep{Yan16b, Wake17}. MaNGA employs dithered observations that contain 17 hexagonal bundles, which offers five different spatial coverages from a diameter of 12 $^{\prime\prime}$ for a 19-fiber bundle to a diameter of 32 $^{\prime\prime}$ for a 127-fiber bundle \citep{Law15}. The raw data are processed by MaNGA Data Reduction Pipeline (DRP), which is described in \citet{Law16}. Our sample is from MaNGA Product Launch-6 (MPL-6).

\subsection{Sample selection}
We first cross-matched 4962 MPL-6 MaNGA sources with the SDSS Data Release 7 (DR7) \citep{Abazajian09} galaxy catalog from MPA-JHU\footnote{\url{https://wwwmpa.mpa-garching.mpg.de/SDSS/DR7/} }, which produced 4320 matches. We use the spectra of the central one spaxel of MaNGA and the single fiber of SDSS-DR7 to select AGN in two databases (Figure \ref{fig:samp} shows the spectra of five examples) with the following criteria:

(1) H$\alpha$ rest equivalent width (EW) are larger than 3\AA, which rejects galaxies with weak emission lines powered by evolved stars \citep[e.g.,][]{Stasinska08,Yan12,Belfiore16,Bing19}. 

(2) The signal-to-noise ratio (S/N) for the emission lines of [NII] $\lambda$6583, H$\alpha$, [OIII] $\lambda$5007 and H$\beta$ are better than 3.0. The emission lines have been measured by Data Analysis Pipeline (DAP) for MaNGA sources \citep{Westfall19}.

(3) The line ratio is located in the AGN region according to the BPT diagram, i.e., [NII] $\lambda$6583/H$\alpha$ versus [OIII] $\lambda$5007/H$\beta$ \citep{Baldwin81, Kauffmann03, Kewley01, Kewley06}. Figure \ref{fig:bpt} shows some examples of BPT diagrams for objects in our sample.

Before further analysis, both the spectra of DR7 and MaNGA have been corrected the galactic reddening using the reddening law of \citet{ODonnell94}. DR7 provides the spectrum of central 3 $^{\prime\prime}$ for each galaxy, so we stack the MaNGA spectrum of central 3 $^{\prime\prime}$ for further analysis. The above criteria yield 195 AGNs in MaNGA, and all of them have been observed by SDSS-DR7. Among 4320 SDSS-DR7 sources that have MaNGA followups, there are 711 galaxies classified as AGNs in the catalog of the SDSS-DR7 database, but only 188 AGNs meet the above 3 selection criteria. In total, we have 231 objects that have two-epoch observations and are classified as AGN during at least one epoch. For these two-epoch spectra (MaNGA and DR7), we normalized the spectra at 5400$\rm\AA$ $\sim$ 5800$\rm\AA$. Then we compared the line width of the normalized spectra of these two-epoch by eye. If the line width of H$\alpha$ or H$\beta$ has significant variations, then we further to measure the line width and determine the exact variations. To determine the line width, we used pPXF package \citep{Cappellari04,Cappellari17} to measure the stellar continuum using MILES Library of Stellar Spectra \citep{Sanchez-Blazquez06,Falcon-Barroso11} and then subtracting a stellar component from each spectrum. After subtracting the stellar component from each spectrum, we fit the emission lines with a single Gaussian component. If this model matched the line profile, and the full width at half maximum (FWHM) was smaller than 500 $\rm km \ s^{-1}$, the feature was classified as a narrow line. If the single Gaussian failed to produce a reasonable match, and additional Gaussian was included in the fit. If the multi-component model fit the profile, and one of the components had a FWHM larger than 2000 $\rm km \ s^{-1}$ and the peak S/N per spectroscopic pixel exceeded 3.0, the line was classified as broad. If an object had a given line change from ``broad'' to ``narrow'' (or the reverse), the source was classified as a CL-AGN. 

This process identified five CL-AGNs. The CL-AGN fraction is about 0.11\% (5 out of 4320 galaxies), which is much higher than the finding (0.007\%) of \citet{Yang18} probably because we focus on nearby objects with high S/N. We also used other line ratio diagnostics ([S II], [O I]) to check these five sources. For the [S II] line ratio diagnostics, two galaxies are not in the AGN region (8322-3701 and 9048-1902). For the [O I] line ratio diagnostics, one galaxy is not in the AGN region (9048-1902). Before further analysis, we used the package of PyQSOFit \citep{Guo18} to decompose the AGN components (e.g. power-law continuum, [Fe II] component) directly. We ran the fitting with and without the AGN continuum, and compared the reduced chi-square. We found that the difference is very small so that the AGN continuum should be negligible. Figure \ref{fig:samp} presents the spectra of two epochs and the best fits for each AGN. Two of them (8940-12702 and 9048-1902) were also followed up with the Hale Telescope (P200), which are further discussed in Section 3. Table \ref{tab:info} lists the basic information for CL-AGN candidates. The rest-frame timescale for the type change is less than 16 years in all cases (Table \ref{tab:info} column (8)). Table \ref{tab:fit} lists the final broad emission lines fitting results for CL-AGNs. The detection significance of the broad Balmer emission lines larger than 5$\sigma$. In order to quantify the reliability of the line change for these sources as CL-AGN, we calculate the flux deviation ($N_{\sigma}(\lambda)$) using the similar way (equation 1) of \citet{MacLeod19}. We define the maximum flux deviation of the line relative to the continuum $N_{\sigma}(4750-4940 \ $\AA$)$ -  $N_{\sigma}(4750 \ $\AA$)$ as $N_{\sigma}(H\beta)$ for 9048-1902, and the line relative to the continuum $N_{\sigma}(6330-6700 \ $\AA$)$ -  $N_{\sigma}(6330 \ $\AA$)$ as $N_{\sigma}(H\alpha)$ for the remaining four sources. As listed in Table \ref{tab:fit}, all of the $N_{\sigma}(H\alpha)$ or $N_{\sigma}(H\beta)$ are larger than 3, meaning that the CL-AGN have reliable disappearance (or appearance) of broad $H\alpha$ or $H\beta$.

\begin{figure*}
\centering
\includegraphics[scale=0.36]{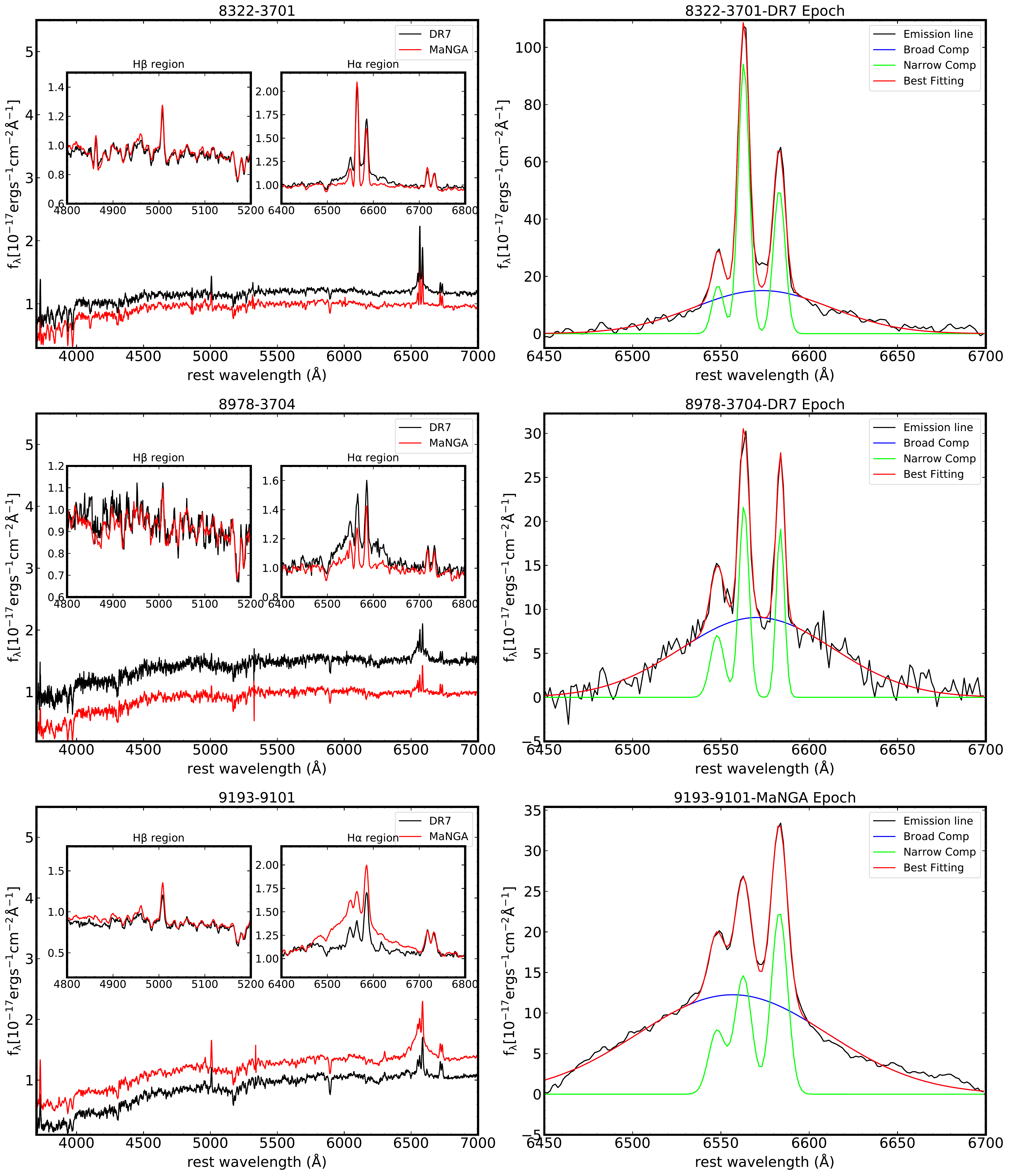}
\caption{Left: Final selected sample of five CL-AGN spectra before subtracting the stellar component within different observation epochs. The black lines indicate the DR7 epoch. The red spectra are the MaNGA data produced by combing the central 3 $^{\prime\prime}$ spectra. The green lines are the latest P200 spectra. The two insets in each panel show the H$\beta$ and H$\alpha$ emission regions. From the DR7 epoch to MaNGA (or P200) epoch, the broad Balmer emission lines disappeared between two objects (``turned off''); the remaining three sources the broad Balmer emission appeared (``turned on''). Right: After subtracting the stellar component, we used two components (broad component and narrow component) to fit the broad H$\alpha$ or H$\beta$ emission lines. For each panel, the title shows the observational epoch. The black lines represent the broad emission lines after subtracting the stellar component. The blue lines and the green lines represent the broad component and the narrow component for the best fit, respectively. The red lines represent the total of the broad and narrow components.}
\label{fig:samp}
\end{figure*}

\begin{figure*}
\setcounter{figure}{0}
\centering
\includegraphics[scale=0.34]{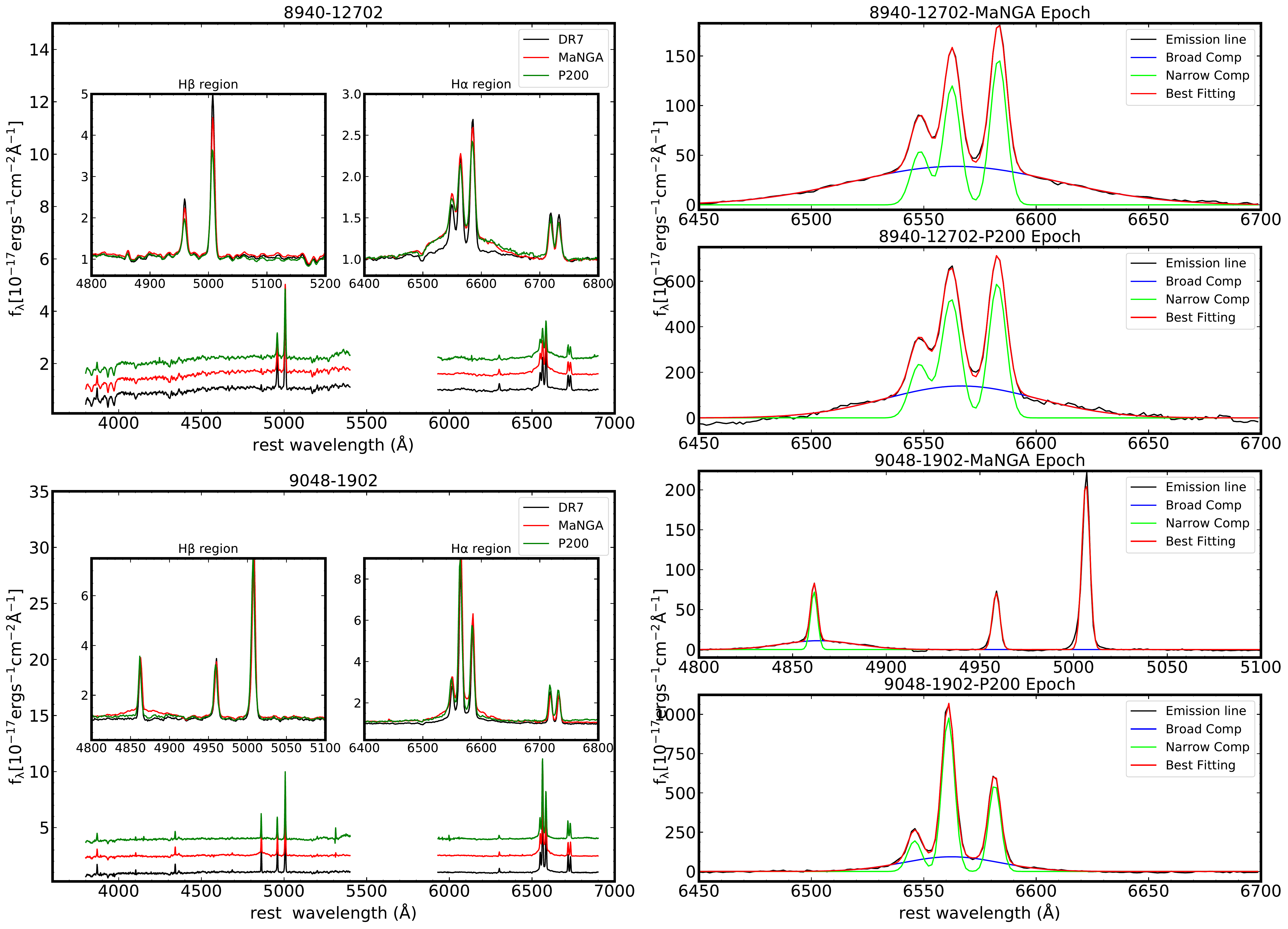}
\caption{(continued)}
\label{fig:cont}
\end{figure*}

\begin{table*}
\centering
\caption{Basic informations of our Changing-look AGN}
\label{tab:info}
\begin{tabular}{cccccccccccc}
    \hline
    SDSS Name & Plate-IFU & RA (J2000) & DEC (J2000) &  $z$  & Epoch-DR7$^{\rm a}$  & Epoch-MaNGA$^{\rm a}$ & $\Delta t_{\rm spec}^{\rm b}$ & state (on/off) \\
    (1) &(2)  & (3)  &(4)  &(5)  &(6)  & (7) & (8)(yr) & (9)\\
    \hline
    
SDSS J131615.95+301552.2 & 8322-3701 & 13:16:15.95 & +30:15:52.27 & 0.0492 & 53852 & $57919$ & 10.62 & off\\
SDSS J080020.98+263648.8 & 8940-12702 & 08:00:20.98 & +26:36:48.81  & 0.0267 & 52581 & $57727$ & 13.73 & on\\
SDSS J163629.66+410222.4 & 8978-3704 & 16:36:29.66 & +41:02:22.45 & 0.0473 & 52379 & $57463$ & 13.30 & off \\
SDSS J162501.43+241547.3 & 9048-1902 & 16:25:01.43 & +24:15:47.3& 0.0503 & 53476 & $57870$ & 11.46 & on \\
SDSS J030510.60-010431.6 & 9193-9101 &03:05:10.60 & -01:04:31.63 & 0.0451 & 51873 & $57693$ & 15.26 & on \\

    \hline
\end{tabular}
  
\begin{tablenotes}
\item (a): Epoch given in modified Julian date
\item (b): $\Delta t_{\rm spec}$ is in rest-frame.
\item (c): ``off'' means from DR7 to MaNGA, broad emission lines disappeared. ``on'' means from DR7 to MaNGA, broad emission lines appeared.

\end{tablenotes}
      
 \end{table*}

\begin{table*}
\centering
\caption {The broad emission line properties of Changing-look AGN}
\label{tab:fit}
\begin{tabular}{cccccccccccc}
    \hline
     Plate-IFU & Flux & $\rm{log_{10}L}$ & FWHM &  Velocity Offset  & Reduced-chi Square  & $N_{\sigma}$ &Broad lines$^{\rm a}$ \\
     &$\rm{(10^{-17} \ erg \ s^{-1} \ cm^{-2}})$  & $\rm{(erg \ s^{-1} })$ &$(\rm{km \ s^{-1}})$  &$(\rm{km \ s^{-1}})$  &  &$H_{\alpha} \ (or \ H\beta)$ & \\
    \hline
    
8322-3701 & $1517.52 \pm 49.00$ & $40.940 \pm 0.014$ &$4483.92 \pm 162.12$ & $510.56 \pm 61.37$ & 0.68 & 6.68 & $H\alpha$ \\
8940-12702 & $4872.35 \pm 21.30$& $40.901 \pm 0.002$ & $4911.07 \pm 25.36$  & $38.50 \pm 8.74$ & 1.69 & 6.45& $H\alpha$ \\
8978-3704 & $1010.46 \pm 34.12$ & $40.728 \pm 0.015$ & $4758.34 \pm 181.06$ & $360.48 \pm 71.33$ & 0.61 & 3.20& $H\alpha$ \\
9048-1902 & $626.91 \pm 7.11$ & $40.576 \pm 0.005$ & $2815.86 \pm 38.13$& $145.46 \pm 15.14$ & 1.37 & 7.18& $H\beta$ \\
9193-9101 & $2050.81 \pm 12.20$ & $40.992 \pm 0.003$ & $5992.49 \pm 41.60$ & $-327.79 \pm 15.27$ & 1.76 & 8.43& $H\alpha$ \\

    \hline
\end{tabular}
  
\begin{tablenotes}
\item (a): $H\alpha$ means that the fitting results from the broad $H\alpha$ emission lines. $H\beta$ means that the fitting results from the broad $H\beta$ emission lines.

\end{tablenotes}
      
 \end{table*}

\section{Data analysis and results}
Figure \ref{fig:samp} shows the observed spectra at these two epochs (DR7 and MaNGA). According to the features of Balmer emission lines ($H\alpha$ or $H\beta$), we found that in two of them the broad $H\alpha$ emission lines disappear. For the other three, two show the presence of strong broad $H\alpha$ emission lines, and the remaining one shows the emergence of strong broad $H\beta$ lines. We observed two of these CL-AGNs (8940-12702 and 9048-1902, both moving from ``off'' to ``on'' states) with Double Spectrograph (DBSP) mounted on the Palomar Hale Telescope (P200) on April 6, 2019. The slit widths we used is 1.5''. The average seeing is 1.5''. The total exposure time for 8940-12702 is 8400s with the average S/N for continuum is about 50. For the 9048-1902, the total exposure time is 9000s with the average S/N for continuum is about 30. The spectral resolution is about 1100 (R = 1100) for the blue band and about 1800 (R = 1800) for the red band. IRAF\footnote{\url{ http://iraf.noao.edu/} } was used to process the original images and extract the spectra. We selected a similar aperture with DR7 to extract the spectra. Figure \ref{fig:samp} (blue lines of left panels) shows the spectra of these two objects. Figure \ref{fig:samp} (right panels) presents the emission line fitting results to these spectra. The rest frame time difference from MaNGA epoch to P200 epoch is about 2 years. The broad Balmer emission lines are still detected, implying that the Type II supernovae (SNe IIn) scenario for these two objects can be ruled out. The third column of Table 2 shows that for the other three sources, the broad $H\alpha$ luminosity is slightly higher than the peak of $H\alpha$ luminosity for most of the type II supernovae (Figure 14 of \citet{Taddia13}). Due to the lack of more observational information for these sources, we can not conclusively rule out their origin in a supernova.

\subsection{The kinematics properties of CL-AGNs}
\begin{figure*}
\includegraphics[scale=0.3]{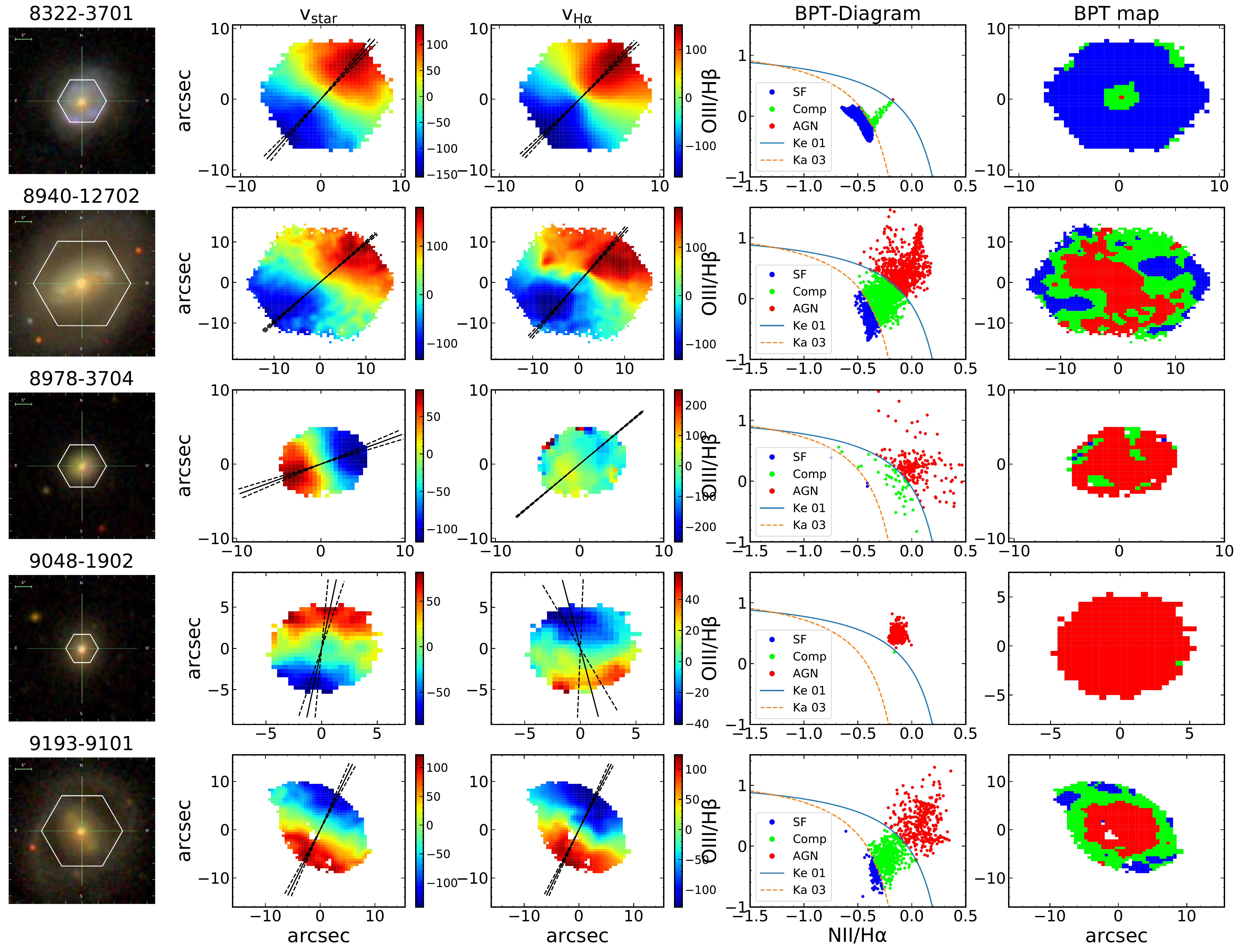}
\caption{MaNGA observations of our sample of five CL-AGNs. The first column presents the SDSS false-colour image; the white hexagon marks the region covered by MaNGA bundle. The second and third columns display the stellar and gas velocity fields, respectively. The solid black line in each velocity field indicates the major axis fitted by FIT\_KINEMATIC\_PA, and two dashed black lines show the 1$\sigma$ error. The last two columns show the BPT diagrams and the BPT diagram maps; the light blue solid curve (``Ke 01“) and the orange dashed curve (``Ka 03'') are the boundary of star-formation, composite and AGN regions from \citet{Kewley01,Kewley06} and \citet{Kauffmann03}. The blue area represents star-formation regions, the green area is the composite region. The remaining region (red) is dominated by AGN. The spatially resolved information (all the flux of emission lines, H$\alpha$ velocity and H$\alpha$ velocity dispersions) are obtained by fitting the emission lines with the MaNGA DAP pipeline \citep{Westfall19}.}
\label{fig:bpt}
\end{figure*}

 We now investigate the gas and stellar kinematic properties of these five CL-AGNs. Figure \ref{fig:bpt} shows the SDSS images, stellar rotation maps, H$\alpha$ rotation maps, and BPT diagrams for all five CL-AGNs. These spatially resolved maps are produced by MaNGA DAP \citep{Westfall19}. The methods of \citet{Krajnovic06} in Appendix C were used to estimate the kinematics position angle (PA). The kinematic misalignment between gas and stars was quantified by $\Delta PA = |PA_{*}-PA_{gas}|$, where $PA_{*}$ and $PA_{gas}$ are the PA of stars and gas, respectively. As listed in Table \ref {tab:2} and Figure \ref{fig:bpt}, one of the five sources show counter-rotating features ($\Delta PA = 160^{\circ}$), while the remaining four display aligned gas and stellar kinematics ($\Delta PA < 30^{\circ}$). \citet{Chen16} found that 9 of 489 blue star-forming galaxies are counter-rotators ($\Delta PA > 150^{\circ}$). The fraction is about $1.84\% \pm 0.61\%$. Our CL-AGNs show a higher but not statistically significant fraction ($20\% \pm 16\%$) of counter-rotating features than normal blue star-forming galaxies.

\begin{table*}
\centering
\caption{The kinematic position angle (PA), black hole mass, stellar velocity dispersion and Eddington ratios}
\label{tab:2}
\begin{tabular}{cccccccccccc}
    \hline
    Plate-IFU & $PA_{*}$ &$PA_{\rm gas}$ & $\Delta PA$ & $M_{\rm BH}$  & $\sigma_{*}$ & $L_{bol}/L_{edd}$\\
     &(deg)& (deg) & (deg)  &($\times 10^{6} \rm M_{\sun}$) & ($\rm km \ s^{-1}$) & \\
    \hline
    
8322-3701 & $141.5 \pm 2.7$ & $138.5 \pm 2.2$ & $3.0 \pm 3.5$ & $11.49^{+2.72}_{-2.26}$ & $132.80 \pm 3.84$ & 0.08\\
8940-12702 & $134.5 \pm 1.2$ & $142.0 \pm 1.7$ & $7.5 \pm 2.1$& $13.26^{+3.02}_{-2.45}$ & $145.80 \pm 5.59$  & 0.06\\
8978-3704 & $112.5 \pm 3.5$ & $133.5 \pm 0.5$ & $21.0 \pm 3.5$& $9.95^{+2.41}_{-2.02}$ & $101.26 \pm 4.09$ & 0.05\\
9048-1902 & $171.0 \pm 5.0$ & $11.0 \pm 12.7$ & $160.0 \pm 13.7$& $4.56^{+0.41}_{-0.41}$ & $103.75 \pm 10.31$ & 0.29\\
9193-9101 &$156.5 \pm 2.5$ & $156.5 \pm 1.7$ & $0.0 \pm 3.1$& $22.44^{+5.21}_{-4.28}$ & $168.47 \pm 4.87$  & 0.03\\

    \hline
\end{tabular}
 \end{table*}

\subsection{Global star formation rate (SFR), bulge kinematics and axial ratio (b/a)}
\begin{figure*}
\includegraphics[scale=0.45]{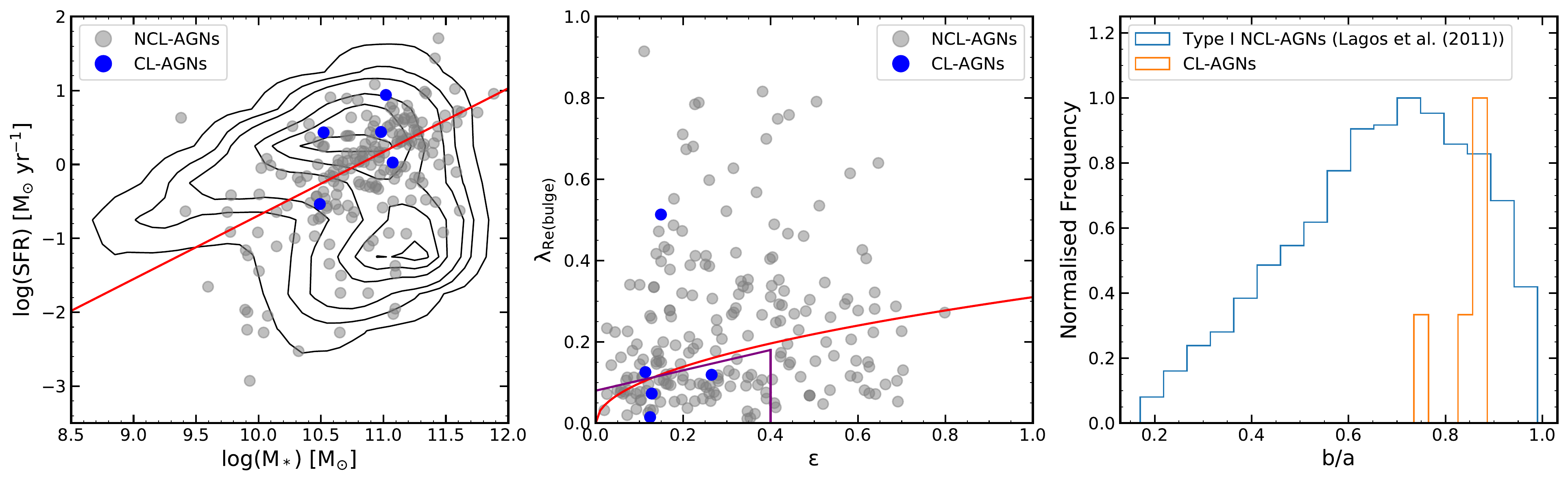}
\caption{Left panel: The SFR-$M_{*}$ diagram. Contours indicate the GSWLC sample. The blue filled circles are the CL-AGNs and the gray color filled circle represent the NCL-AGNs classified by us using MaNGA MPL-6 sources. The red dashed line is an approximation of the boundary of the star-forming main sequence, which is adopted the fig. 11 of \citet {Chang15} at the 1$\sigma$ level in scatter. The approximate slope and intercept are from \citet{Jin16}. Middle panel: The $\lambda_{Re (bluge)}$-$\epsilon$ diagram. The blue filled circles indicate the CL-AGNs and the gray color filled circle represent the NCL-AGNs. The solid red line is the division between fast and slow rotators proposed by \citet{Emsellem11}; the purple line is the fast/slow rotator separation proposed by \citet{Cappellari16}. Objects above of these lines are ``fast rotators'' and at the bottom of these lines means slow rotator. The right panel displays the distribution of b/a for CL-AGNs and Type I NCL-AGNs. The Type I NCL-AGNs are from Lagos et al. (2011).}
\label{fig:main}
\end{figure*}

We cross-matched our CL-AGN candidates with the GALEX-SDSS-WISE Legacy Catalogue (GSWLC) \citep{Salim16, Salim18} to obtain global SFRs and stellar masses ($M_{*}$). In order to compare the properties of CL-AGNs and non-changing-look AGNs (NCL-AGNs), we defined 226 NCL-AGNs, which do not show any type of change between MaNGA and DR7 epochs. A caveat here is that by comparing only two spectra that are $\sim$10 years apart it is  possible that some sources did not change the type and back again in between the two observations. The NCL-AGNs sample may therefore contain some CL-AGNs which were not caught. Additionally, some CL-AGNs may not be identified as such because the data quality is not sufficiently good to detect the line change. But all of these outliers should be a tiny fraction, which should not affect our results. The left panel of Figure \ref{fig:main} displays the location of our CL-AGNs candidates along with normal AGNs on the SFR-$M_{*}$ plane. The red solid line is the boundary of the star-forming main sequence. \citet{Charlton19} used the way of S\'ersic indices versus colors diagram to separate galaxies as blue disks, green valley and red spheroids, and found that all of their faded CL-AGNs host galaxies located in the green valley region. The possible reason is that we defined the star-forming main sequence in different ways. We also used SFR versus stellar mass to check the faded CL-AGNs host galaxies of \citet{Charlton19}. Although there is only one target of \citet{Charlton19} that has been matched with MPA-JHU catalog, this galaxy is located in the star-forming main sequence region. If all of our 5 CL-AGNs used SFRs from MPA-JHU catalog, they still locate in the star-forming main sequence region. This panel demonstrates that most of CL-AGNs and NCL-AGNs located within the star-forming main sequence region. There are no significant differences between CL-AGNs and NCL-AGNs.

The middle panel of Figure \ref{fig:main} shows the bulge kinematics for CL-AGNs and NCL-AGNs. For early-type galaxies observed by IFU, \citet{Emsellem07} and \citet{Cappellari16} provided a simple parameter $\lambda_{R}$ to trace the stellar angular momentum of a galaxy
\begin{equation}
\lambda_{R}  \equiv \frac {\langle R|V|\rangle}{\langle R\sqrt{V^{2}+\sigma^{2}}\rangle} = \frac {\sum_{n=1}^N F_{n}R_{n}|V_n|}{\sum_{n=1}^N F_{n}R_{n}\sqrt{V^{2}+\sigma^{2}}}
\end{equation}
Where F is the flux, R is the circular radius, V is the stellar velocity, $\sigma$ is the stellar velocity dispersion and N is the number of spaxels. A large $\lambda_{R}$ represents a fast rotator, and small $\lambda_{R}$ indicates a slow rotator. The SFR versus $M_{*}$ diagram (left panel of Figure \ref{fig:main}) reveals that most of the CL-ANGs and NCL-AGNs are star-forming galaxies (late type). We applied this same technique to measure $\lambda_{Re}$ of bulges ($\lambda_{Re (bluge)}$) of our CL-AGNs that have spiral morphologies. \citet{Emsellem11} provided an empirical diagnostic to distinguish fast and slow rotators according to the diagram of $\lambda_{Re}$ versus $\epsilon$, where $\epsilon$ is the apparent ellipticity ($\epsilon = 1 - b/a$). The middle panel of Figure \ref{fig:main} shows the  $\lambda_{Re (bluge)}$-$\epsilon$ diagram for CL-AGNs and other NCL-AGNs. The bulge Re is measured by MaNGA PyMorph Photometric Value Added Catalogue (MPP-VAC) \citep{Fischer19}. The middle panel of Figure \ref{fig:main} shows three CL-AGNs possess slow-rotator bulges, meaning that galaxies may have gone the dry merger process. The other two CL-AGNs have fast-rotator bulges, implying that galaxies have undergone the gas-rich merging or gas accretion processes \citep{Cappellari16}. NCL-AGNs and CL-ANGs have similar distributions in the $\lambda_{Re (bluge)}$-$\epsilon$ diagram.

In order to compare the properties of b/a between CL-AGNs and normoal AGNs, we construct a sample of 6070 Type I NCL-AGNs from \citet{Lagos11}. The right panel in Figure \ref{fig:main} presents the distribution of b/a for our CL-ANGs and their Type I NCL-AGNs. The b/a of CL-AGNs are drawn from the data release by MaNGA\footnote{\url{https://dr15.sdss.org/sas/dr15/manga/spectro/analysis/v2_4_3/2.2.1/}} DAP \citep{Westfall19}. The b/a of Type I NCL-AGNs are drawn from \citet{Lagos11}. This panel indicates that all of our CL-AGNs are face-on galaxies, with b/a $>$ 0.7. We randomly select 5 sources from the Type I NCL-AGNs and found that there is a 2.28\% probability that the selected galaxies have b/a $>$ 0.7. It appears that CL-AGNs favor more face-on than that of Type I NCL-AGNs. These implies that the host galaxy obscuration may hamper the discovery of CL-AGNs.

\begin{figure*}
\centering
\includegraphics[scale=0.7]{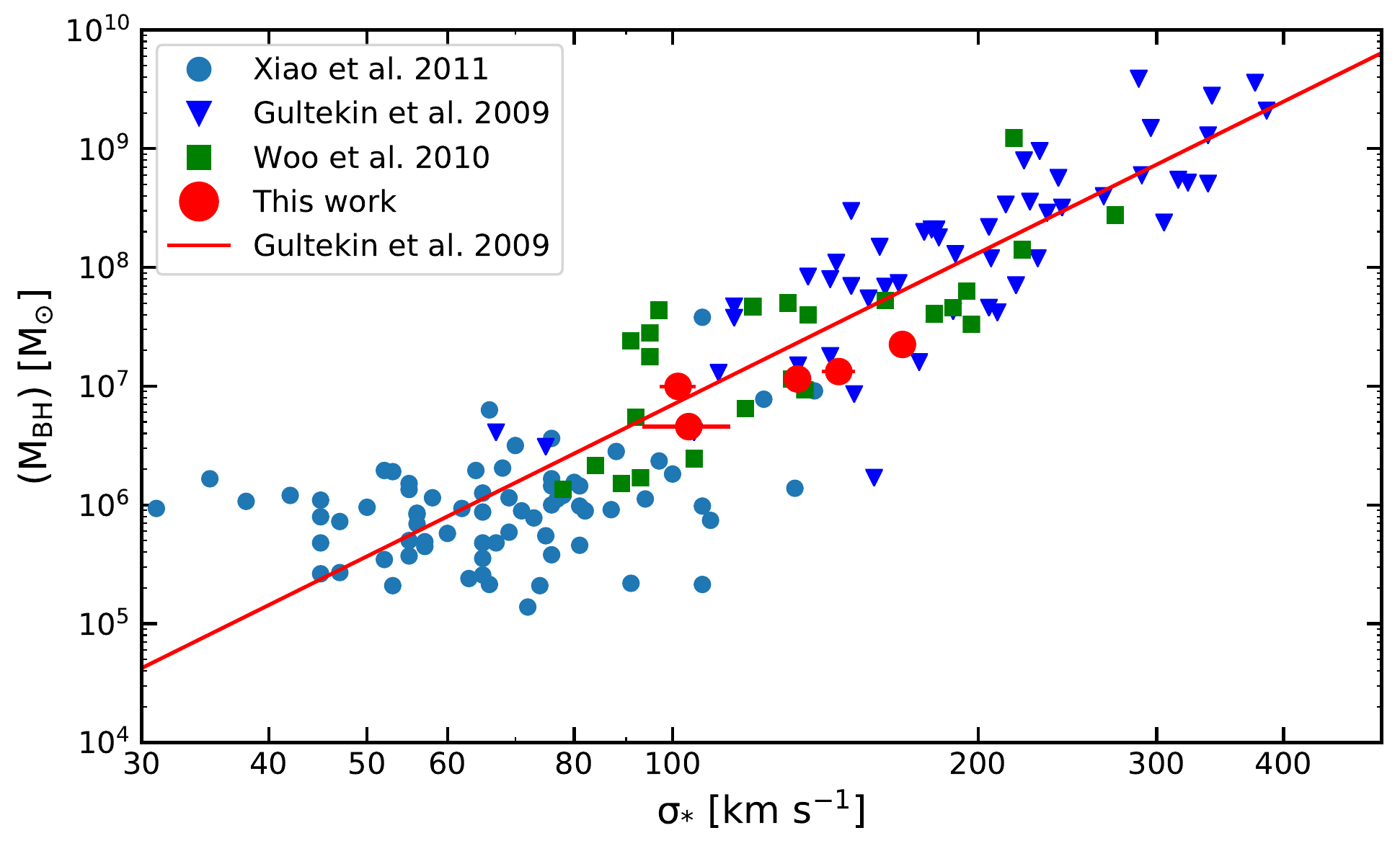}
\caption{$M_{BH}-\sigma_{*}$ relation diagram for CL-AGNs and a set of normal AGNs. The red filled circles represent our sample of five CL-AGNs. other symbols represent normal AGNs from \citet{Gultekin09,Xiao11} and \citet{Woo10}. The red solid line indicates the $M_{BH}-\sigma_{*}$ relation from  \citet{Gultekin09}.}
\label{fig:m-sigma}
\end{figure*}

\subsection{Bulge properties}

The S\'ersic index is a useful parameter to classify the morphologies of our CL-AGNs. We use GALFIT \citep{Peng02,Peng10} to fit the surface brightness profile for SDSS g, r and i band images. \citet{Charlton19} studied four faded CL-AGNs using broadband images. They pointed out that although the quasars have faded, usually they didn't disappear completely. So when they fitted the surface brightness profile, they added a PSF-like component to avoid central concentrations. Following the theory of \citet{Charlton19}, when we use GALFIT to fit the surface brightness profile for our CL-AGNs, we used both one S\'ersic model and two S\'ersic models to fit them. The fitting results are listed in the Table \ref{tab:sersic}. Two components fitting are needed for four CL-AGNs while only one component is enough for the remaining. Table \ref{tab:sersic} shows that four galaxies have pseudo-bulge (S\'ersic index $<$ 4) and one has a classical bulge (S\'ersic index $>$ 4) features.

\begin{table*}
\centering
\caption{S\'ersic index fitting results from GALFIT}
\label{tab:sersic}
\begin{tabular}{cccccccccccc}
    \hline
    Plate-IFU & Band &PSF & Component Lable & S\'ersic index  & Reduced-chi Square \\
     & & (Magnitude) &   & & \\
    \hline
    
8322-3701 & g & $17.90 \pm 0.01$ & S\'ersic 1 & $1.51 \pm 0.02$ & 1.43  \\
          &   &                  & S\'ersic 2 & $0.06 \pm 0.01$ &       \\
          & r & $17.21 \pm 0.01$ & S\'ersic 1 & $2.12 \pm 0.02$ & 1.28  \\
          &   &                  & S\'ersic 2 & $0.06 \pm 0.01$ &       \\
          & i & $16.81 \pm 0.01$ & S\'ersic 1 & $2.28 \pm 0.03$ & 1.26  \\
          &   &                  & S\'ersic 2 & $0.05 \pm 0.01$ &       \\

8940-12702& g & $18.18 \pm 0.01$ & S\'ersic 1 & $5.16 \pm 0.06$ & 1.37  \\
          &   &                  & S\'ersic 2 & $0.32 \pm 0.01$ &       \\
          & r & $18.68 \pm 0.02$ & S\'ersic 1 & $7.47 \pm 0.09$ & 1.44  \\
          &   &                  & S\'ersic 2 & $0.48 \pm 0.01$ &       \\
          & i & $18.30 \pm 0.03$ & S\'ersic 1 & $9.08 \pm 0.12$ & 1.33  \\
          &   &                  & S\'ersic 2 & $0.49 \pm 0.01$ &       \\

8978-3704 & g & $20.29 \pm 0.08$ & S\'ersic 1 & $1.47 \pm 0.02$ & 1.21  \\
          & r & $18.55 \pm 0.03$ & S\'ersic 1 & $1.52 \pm 0.01$ & 1.22  \\
          & i & $18.52 \pm 0.01$ & S\'ersic 1 & $1.47 \pm 0.01$ & 1.19  \\

9048-1902 & g & $19.56 \pm 0.07$ & S\'ersic 1 & $1.18 \pm 0.05$ & 1.17  \\
          &   &                  & S\'ersic 2 & $1.78 \pm 0.09$ &       \\
          & r & $18.61 \pm 0.07$ & S\'ersic 1 & $1.79 \pm 0.06$ & 1.16  \\
          &   &                  & S\'ersic 2 & $0.62 \pm 0.08$ &       \\
          & i & $17.82 \pm 0.06$ & S\'ersic 1 & $2.25 \pm 0.03$ & 1.14  \\
          &   &                  & S\'ersic 2 & $0.53 \pm 0.08$ &       \\

9193-9101 & g & $18.34 \pm 0.01$ & S\'ersic 1 & $1.77 \pm 0.04$ & 1.13  \\
          &   &                  & S\'ersic 2 & $0.15 \pm 0.01$ &       \\
          & r & $17.74 \pm 0.01$ & S\'ersic 1 & $1.88 \pm 0.02$ & 1.22  \\
          &   &                  & S\'ersic 2 & $0.16 \pm 0.01$ &       \\
          & i & $17.48 \pm 0.01$ & S\'ersic 1 & $2.55 \pm 0.03$ & 1.28  \\
          &   &                  & S\'ersic 2 & $0.16 \pm 0.01$ &       \\

    \hline
\end{tabular}
  
\begin{tablenotes}
\item Note: The third column shows the magnitude of the PSF component added to each fit. The fifth column shows the fitting results of the S\'ersic index. The last column shows the goodness of each fit.
\end{tablenotes}
      
 \end{table*}

\subsection{Black Hole Mass ($M_{BH}$) and $M_{BH} - \sigma$ relation}
CL-AGNs offer an opportunity to study the $M_{BH} - \sigma$ relationship: when the broad lines are present the black-hole mass can be measured; when the broad lines disappear and the AGN continua fades, the velocity dispersion of bulges can be measured. We used the formula of \citet{Greene05} to estimate the black-hole masses from the line width of broad H$\alpha$ and H$\beta$. Table \ref{tab:fit} listed the fitting results of broad Balmer emission lines. The stellar velocity dispersions ($\sigma_{*}$) and the errors are also provided by pPXF. The instrument velocity dispersion has been removed \citep{Westfall19}. The derived $M_{BH}$ and $\sigma_{*}$ are listed in Table \ref{tab:2}.

Figure \ref{fig:m-sigma} shows the distribution of CL-AGNs in the $M_{BH}-\sigma_{*}$ plane with NCL-AGNs. As shown in the figure, CL-AGNs generally follow the same trend as the NCL-AGNs. In Section 3.3 it was shown that 80\% of our CL-AGNs possess pseudo-bulge features. \citet{Ho14} reported that the relation for pseudo-bulges has a different zero point and much larger scatter than classical bulges and ellipticals in $M_{BH} - \sigma$ relation. Unfortunately in our work the number of CL-AGNs is too small to make a robust determination of the relationship itself.

Some recent results pointed out that CL-AGNs have systematically lower Eddington ratios than NCL-AGNs \citep[e.g.,][]{Rumbaugh18,MacLeod19}. \citet{Greene05} provided the empirical correlations between broad H$\alpha$ (or H$\beta$) luminosities and optical continuum luminosity ($L_{5100}$). According to the bolometric correction factors provided by \citet{Netzer19}, we estimated the bolometric luminosities for each CL-AGNs by $L_{5100}$. Then we estimated the Eddington ratio ($L_{bol}/L_{edd}$) for each source, where $L_{bol}$ is the bolometric luminosity and $L_{edd}$ is the Eddington luminosity, as listed in the last column of Table \ref{tab:2}. Compared the Eddington ratios with \citet{Rumbaugh18} and \citet{MacLeod19}, the results show that the CL-AGNs have systematically lower Eddington ratios than that of NCL-AGNs.

\section{summary}
In this work, we identified 5 CL-AGNs by combining data from SDSS-I/II (DR7) and MaNGA, and studied the host galaxy properties for these sources. The main results are as follows:

(1) Most of CL-AGNs and NCL-AGNs located within the star-forming main sequence region.

(2) Our CL-AGNs show a higher but not statistically significant fraction ($20\% \pm 16\%$) of counter-rotating features as compared to that ($1.84\% \pm 0.61\%$) in general blue star-forming galaxies.

(3) CL-AGNs favor more face-on than that of Type I NCL-AGNs, implying that the host galaxy obscuration may hamper the discovery of CL-AGNs.

(4) CL-AGNs and NCL-AGNs have similar distributions in the $\lambda_{Re (bluge)}$-$\epsilon$ diagram.

(5) The $80\% \pm 16\%$ of our CL-AGNs do possess pseudo-bulge features, and follow the overall NCL-AGNs $M_{BH}-\sigma_{*}$ relationship.

(6) According to the latest observation spectra and the type transition timescales, the transition mechanisms of SNe IIn scenario has been excluded for two of our CL-AGNs.

Overall, The host galaxy properties give some new insights to comprehend some special properties of CL-AGNs. Although some researches pointed out that the changing-look phenomenon is related to rapid changes in the inner accretion disc \citep[e.g.,][]{Noda18,Ross18}. The larger-scale environment of CL-AGNs can provide the environment where the black hole grows and more insights into the circumstances influencing co-evolution.

\section{acknowledgements}
We thank the referee for a detailed report that helped significantly in improving the presentation of our work. X.Y. and Y.S. acknowledge the support from the National Key R\&D Program of China (No. 2018YFA0404502, No. 2017YFA0402704), the National Natural Science Foundation of China (NSFC grants 11825302, 11733002 and 11773013). Y. S. thanks the support from the Tencent Foundation through the XPLORER PRIZE. R.A.R thanks partial financial support from Conselho Nacional de Desenvolvimento Cient\'ifico e Tecnol\'ogico (202582/2018-3 and 302280/2019-7) and Funda\c c\~ao de Amparo \`a pesquisa do Estado do Rio Grande do Sul (17/2551-0001144-9 and 16/2551-0000251-7). RR thanks CNPq, Capes, and FAPERGS for financial
support. The authors thank Jujia Zhang for his help of P200 data reduction. The  authors thank Donald Schneider's valuable suggestions.

This research uses data obtained through the Telescope Access Program (TAP), which has been funded by the National Astronomical Observatories, Chinese Academy of Sciences, and the Special Fund for Astronomy from the Ministry of Finance. Observations obtained with the Hale Telescope at Palomar Observatory were obtained as part of an agreement between the National Astronomical Observatories, Chinese Academy of Sciences, and the California Institute of Technology.

Funding for the Sloan Digital Sky Survey IV has been provided by the Alfred 
P. Sloan Foundation, the U.S. Department of Energy Office of Science, and the 
Participating Institutions. SDSS-IV acknowledges
support and resources from the Center for High-Performance Computing at
the University of Utah. The SDSS web site is www.sdss.org.

SDSS-IV is managed by the Astrophysical Research Consortium for the 
Participating Institutions of the SDSS Collaboration including the 
Brazilian Participation Group, the Carnegie Institution for Science, 
Carnegie Mellon University, the Chilean Participation Group, the French 
Participation Group, Harvard-Smithsonian Center for Astrophysics, 
Instituto de Astrof\'isica de Canarias, The Johns Hopkins University, 
Kavli Institute for the Physics and Mathematics of the Universe (IPMU) / 
University of Tokyo, Lawrence Berkeley National Laboratory, 
Leibniz Institut f\"ur Astrophysik Potsdam (AIP),  
Max-Planck-Institut f\"ur Astronomie (MPIA Heidelberg), 
Max-Planck-Institut f\"ur Astrophysik (MPA Garching), 
Max-Planck-Institut f\"ur Extraterrestrische Physik (MPE), 
National Astronomical Observatories of China, New Mexico State University, 
New York University, University of Notre Dame, 
Observat\'ario Nacional / MCTI, The Ohio State University, 
Pennsylvania State University, Shanghai Astronomical Observatory, 
United Kingdom Participation Group,
Universidad Nacional Aut\'onoma de M\'exico, University of Arizona, 
University of Colorado Boulder, University of Oxford, University of Portsmouth, 
University of Utah, University of Virginia, University of Washington, University of Wisconsin, 
Vanderbilt University, and Yale University.

\section{DATA AVAILABILITY}
The data underlying this article will be shared on reasonable request to the corresponding author.

\bibliography{reference}



%
%


\bsp	
\label{lastpage}
\end{document}